\documentclass[reprint,aps,amsmath,amssymb,]{revtex4-2} %new
\usepackage{graphicx}% Include figure files %new
\usepackage{dcolumn}% Align table columns on decimal point %new
\usepackage{bm}% bold math %new
\begin{document}

%\preprint{APS/123-QED} %new

\title{Finding, mapping and classifying optimal protocols for two-qubit entangling gates} %based on Rydberg blockade}
%entangling gates  on atomic qubits}

\author{Ignacio R. Sola}
\email{corresponding author: isolarei@ucm.es} 
\affiliation{Departamento de Quimica Fisica, Universidad Complutense, 28040 Madrid, Spain}%Lines break automatically or can be forced with \\
  % \email{isolarei@ucm.es}
\author{Seokmin Shin}
 \affiliation{School of Chemistry, Seoul National University, 08826 Seoul, Republic of Korea}
\author{Bo Y. Chang}
 \email{corresponding author: boyoung@snu.ac.kr}
 \affiliation{School of Chemistry, Seoul National University, 08826 Seoul, Republic of Korea}
 \affiliation{Research Institute of Basic Sciences, Seoul National University, 08826 Seoul, Republic of Korea}

\begin{abstract}
We characterize the set of %landscape of 
optimal protocols for two-qubit entangling gates %the CZ gate
%in $N$ qubit systems 
through a mechanism analysis based on quantum pathways,
which allows us to compare and rank the different solutions.
As an example of a flexible platform with a rich landscape of protocols,
we consider %non-independent 
trapped neutral atoms excited to Rydberg
states by different pulse sequences that extend over several atomic sites,
optimizing both the temporal and the spatial features of the pulses.
%Our optimization procedure considers both the temporal and spatial features of the pulses acting on arrays of trapped neutral atoms, allowing for a denser array of atoms, which can potentially speed up the gate operation by two orders of magnitude. 
Studying the rate of success of the algorithm under different constraints,
we analyze the %minimal energy requirements for the gate operation and
%the different participation of the qubits and
impact of the proximity of the atoms on the nature and quality of the optimal protocols.
%, which, indirectly, reveals the role of the geometry of the arrays in $3$ and $4$ qubit systems.
We characterize in detail the features of the solutions in parameter space, 
showing some striking correlations among the set of parameters. 
Together with the mechanism analysis, the spatio-temporal control
allows us to select protocols that operate under mechanisms by design,
like finding needles in the haystack.
\end{abstract}

\maketitle

\section{Introduction}

There are several well-studied platforms 
to build quantum computer prototypes \cite{Cirac_Nature2000,Ladd_Nature2010,Schoelkopf_Science2013,Kelly_Science2015,Harty_PRL2014,Wrachtrup_PRL2004,Saffman_RMP2010},
each with many possible designs proposed to implement different quantum gates.
Their merits are compared with regards to the fidelities achieved, the number of operations that can be executed coherently, and scalability properties.
%\textcolor{red}{Many designs aim to implement various quantum gates with different merits, such as high fidelities, the number of operations that can be executed coherently, and scalability properties.}
%and many designs to implement different quantum gates 
%whose merits dwell in
%the fidelities achieved, the number of operations that can be executed coherently, 
%and scalability properties.
%
Cross-platform comparisons of different quantum computers are starting to emerge 
based on their performance under specific algorithms \cite{Zhu_NatComm2022}.
Almost all the protocols proposed so far were developed through ingenious ideas and
further fine-tuned by numerical and experimental studies.
However, these protocols clearly do not encompass %exhaust 
the number of possible solutions.
It is the main goal of this work to organize, classify, rank, 
and also to visualize %in the simplest way, 
all the possible protocols that can be found for a certain class of
entangling gates given some constraints, 
to serve as a guiding search for promising experimental implementations.

%We seek to answer questions like: What is the minimal energy input introduced through external fields necessary for the gates to operate? 
%And how much per qubit?
%Are all qubits used equally during the gate dynamics?
%How much freedom is there in the pulse parameters (pulse areas, spatial profiles of the laser beams) of the optimal protocols under certain constraints? Are there strict correlations, which must be fulfilled by the parameters revealing some aspects of the dynamics?
%And finally, can we infer which qubit arrangements or geometries are more promising for high fidelity gates? 

To explore the landscape of all possible protocols 
we use the techniques of quantum control.
Quantum control was previously used to find the pulse areas and the sequence of pulses that maximizes the probability of reaching a specific quantum state \cite{Rice_2000,Shapiro_2011,shore_2011,Malinovsky_PRA2004,Malinovsky_PRL2004,Malinovsky_PRL2006},
or a set of states necessary for %implied in 
the realization of a quantum gate \cite{Kosloff_PRA2003,Tesch_CPL2001,Tesch_PRL2002,Koch_PRA2008,Goerz_NJP2014,Caneva_PRA2011,Glaser_EPJD2015,Koch_EPJQT2022,Koch_PRA2011,Saffman_PRA2016}.
Unlike in previous approaches where a specific realization of the gate is imposed, %Instead of forcing a specific realization of the gate, 
here we use quantum
optimal control techniques to scan and characterize the full space of optimal solutions,
working with sequences with different numbers of pulses and features \cite{Rabitz_Science2004,Rabitz_PRA2006,Rabitz_IRPC2007,Christiane_PRA2013,Tannor_PRL2011,Rabitz_PRA2009,Rabitz_JCP2012,Rabitz_JPA2017}.

When the number of parameters to be optimized %changes
is by itself a variable, many alternatives exist
on how to compare and classify the solutions. 
To catalog the different protocols,
we use a mechanistic analysis of the internal operation of the gate, %the inner workings of the gate, 
based on quantum pathways tracking %on 
the set of computational basis and ancillary states visited during the gate dynamics. 
As a step further, we can guide the optimization
algorithm to find an optimal protocol that works by design.

While our approach is general, 
we focus on optimal protocols for entangling gates such as %of the kind of controlled-Z (CZ),
%in this work we find and characterize optimal protocols for entangling
%gates of the kind of controlled-Z (CZ) %and $3$-qubit generalizations, 
controlled-Z(CZ) gates implemented 
%on a specific platform, that should be as flexible as possible to provide a rich landscape of solutions.
%We focus 
on neutral atoms trapped by optical tweezers \cite{Browaeys_PRX2014,Browaeys_Science2016,Wilson_PRL2022,Thompson_PRXQ2022,Ahn_OptE2016}.
These are easily addressable by optical methods and can be entangled through Rydberg blockade \cite{Comparat_JOSAB2010,Tong_PRL2004,Saffman_NPhys2009,Grangier_Nphys2009,Adams_PRL2010},
offering promising applications in preparing multi-particle entanglement \cite{Lukin_PRL2018,Zhan_PRL2017,Ahn_PRL2020,Grangier_PRL2010,Saffman_Nature2022,Saffman_PRA2015,Saffman_PRA2010,Picken_QCT2018,Malinovsky_PRA2004,Malinovsky_PRL2004,Malinovsky_PRL2006} and simple quantum circuits  \cite{Jaksch_PRL2000,Saffman_QIP2011,Saffman_PRA2015,Lukin_PRL2001,Lukin_PRL2019,Cohen_PRXQ2021,Shi_QSTech2022,Shi_PRApp2018,Adams_PRL2014,Adams_JPB2019,Malinovsky_PRA2014,Goerz_JPB2011,Morgado_AVSQSci2021,Alexey_PRL2021,Sanders_PRA2020}. 
In the usual set-up, each qubit is addressed by different lasers independently of the others, 
for which the atoms must occupy largely separated positions in the trap.
As the interaction energy between the atoms becomes much weaker, of the order of the MHz,
the necessary time for the two-qubit gate to operate reaches the microsecond regime.
%One of the problems in the usual designs with ordered arrays of atoms largely separated in optical traps 
%\textcolor{red}{to avoid exciting more than one atom in a Rydberg state using the dipole blockade mechanism} is the long duration of the two-qubit gates, which must operate in the micro- or sub-microsecond regime. 
%to avoid exciting more than one atom in a Rydberg state using the dipole blockade mechanism. 
To speed up the gate, in this work we w will use denser arrays of trapped atoms, 
that allow to boost the dipole-blockade energy near the GHz \cite{Ohmori_NP2022,Alexey_PRL2021}.
The price to pay is that the qubits can no longer be regarded as independent, 
as the laser beams may overlap significantly with more than one qubit site.
%We call this the ``strong dipole-blockade regime'' \cite{Ohmori_NP2022,Alexey_PRL2021}. 
The interrelation of the qubits driven by the fields
%This 
can be regarded as a problem or as an opportunity. 
%since 
By controlling the position of the atoms with respect to the different laser beams, 
and adding a spatial control knob to the problem, 
one gains a novel and important feature
that provides both flexibility and robustness to the gate protocols, in addition to the speed-up.
We will show that trapped atoms with strong dipole blockades %regime 
provide a platform with a rich landscape of optimal protocols.

In a recent contribution \cite{Sola_Nanoscale2023}, we proposed an extension of the CZ gate
protocol of Jaksch et al.\cite{Jaksch_PRL2000} for non-independent two-qubit systems, 
named SOP (symmetric orthogonal protocol), which implied controlling
both the temporal (pulse areas) and spatial properties of the light.
The gate mechanism relied on %exploited 
the presence of a dark state in the Hamiltonian, for
which the pulses in the sequence had to be {\em spatially orthogonal}, 
in the sense that the parameters of these fields at each qubit location
formed orthogonal vectors \cite{Sola_Nanoscale2023}.
In \cite{Sola_Nanoscale2023} we proposed the use of 
hybrid modes of light to force the orthogonality. 
%A possible generalization for spatially non-orthogonal pulses may require more complex structured light\cite{Murty_AppOpt1964}, such as those sketched in Fig.\ref{scheme}. But the spatial control can be achieved by different means, and a simpler laboratory implementation shown in Fig.\ref{scheme}, can be achieved using a superposition of overlapping phase-locked Gaussian modes centered at different qubits,  instead of a single field, for each pulse in the sequence.
%In any case, it is necessary to dispose of sufficiently complex light structures to extend the control to the spatial domain \cite{Forbes_NPhoto2021,Rubinsztein_Dunlop_JOpt2016}. 
In ideal conditions, the set of parameters under which the SOP has maximum fidelity
defines a lattice in parameter space, where the implementation of the
gate is robust, but typically with a relatively low yield ($F \sim 0.98$).
A second goal of this work is to extend the SOP scheme 
by exploring how much some
of its requirements are necessary.
To fully optimize the gate performance in this setup, we have developed optimization
techniques that deal not only with the temporal parameters of the laser but also with the spatial structure of the field.
Our results indicate %It turns out 
that by loosing the very strict restrictions of the SOP, one
can find a rich family of optimal protocols with higher yields.
Depending on the number of pulses in the sequence 
or the operating mechanism of the gate,  
%, or the mechanism by which the gate operates, 
striking correlations in the pulse parameters are found.
Typically, non-obvious correlations in control parameters reveal interesting
structures in the Hamiltonian that are exploited in the dynamics,
a subject for future studies.

%As a first approach on the subject of full quantum control by addressing both the spatial and temporal features of the lasers we will assume ideal conditions (a simplified Hamiltonian, zero temperature, no laser noise).
%Working with fast gates justifies neglecting most sources of decoherence as a first approximation. Preliminary calculations show a relatively weak dependence  of the fidelity on intensity fluctuations in the laser fields, as well as the atomic motion of the atoms, which can lead to a $\sim 1$\% loss of fidelity at $\sim25 \mu$K.
%In addition, the control may be extended to other variables, including the time-delays and the relative phases between the pulses \cite{Kosloff_PRA2003,Tesch_CPL2001,Tesch_PRL2002,Koch_PRA2008,Goerz_NJP2014,Caneva_PRA2011,Glaser_EPJD2015,Koch_EPJQT2022,Koch_PRA2011,Saffman_PRA2016}.
%More detailed studies will be needed to address the practical implementation of the protocols and possible extensions.

%\section{Analytical Model}

\section{Qubit set-up}

%\begin{figure}
%\includegraphics[width=8.5cm]{fig1_new.png} %{scheme-3q-c.png}
%\caption{Scheme showing two possible implementations of the spatio-temporal control of trapped qubits, located at the $\times$ positions, using structured pulses $\Omega_k$, whose spatial profile at the qubits gives the desired local amplitudes (second row), or a linear combination of phase-locked Gaussian beams TEM$_{00}$   $\Omega_a, \Omega_b, \Omega_c$, centered at each qubit, that give the same local fields (third row). The sequence of operations that governs the temporal evolution of the state depends on the pulse sequence, shown in the first row.}
%\label{scheme}
%\end{figure}
 
In neutral atoms \cite{Adams_book2018,gallagher_1994}, 
the computational basis is typically encoded in low-energy
hyper-fined states of the atom.
The C-PHASE implies population return to the initial state with a phase
change conditional on the state of the control qubit.
When the phase is $\pi$, the gate is usually called CZ gate. 
In most protocols, this is achieved with an ancillary state, by driving the population
through a Rydberg state of the atom 
$|r\rangle$, gaining a phase accumulation (for resonant $2\pi$ pulses) of $\pi$.
The pulse frequencies are tuned to excite the chosen Rydberg state %, $|r\rangle$
from the ground state (alternatively, from the $|1\rangle$ state) so the other
qubit state is decoupled.
Doubly-excited Rydberg states cannot be further populated by ladder climbing
%\textcolor{red}{Ladder climbing} %by  exciting further Rydberg states is forbidden or greatly reduced 
due to the dipole blockade mechanism if the atoms are within the radius
blockade distance ($r_{\cal B}$) \cite{Urban_NP2009,Tong_PRL2004}.

When the atoms are sufficiently separated, one can address them independently,
as in the well-known protocol
proposed by Jaksch and collaborators \cite{Jaksch_PRL2000}, %that
which uses the pulse sequence:
$\pi_1 - 2\pi_2 - \pi_1$. %where
In this sequence, the first and last pulses act 
on the first qubit (qubit $A$), and the middle pulse acts of the second qubit
(qubit $B$).
JP demands slow gates because the 
largely separated atoms lead to 
weak dipole blockades $d_{\cal B}$, in the MHz.
However, working with atoms at closer interatomic distances ($d \sim 1 \mu$m) %-2 \mu$m)
one can typically increase the dipole-dipole interaction to almost a GHz,
depending on the atom and the Rydberg state, potentially allowing  to operate
the gate in the  nanosecond regime. %\cite{Sola_Nanoscale2023}.
%With strong magnetic fields, the energy splitting between the qubit states can reach almost $\Delta \sim 10$ GHz \cite{Grangier_PRL2010,Scotto_phd2016}, while the energy difference between adjacent Rydberg states when the principal quantum number is below $n = 80$ is typically larger \cite{gallagher_1994}.     
%Hence, $\Delta^{-1}$ could determine the ultimate limit to operate the gate, in the GHz,  but in the JP the gate time is limited by $d_{\cal B}^{-1}$. The gap in the time-scales  offers an opportunity to speed up the gates typically by two orders of magnitude. This requires using denser arrays of atoms, with inter-atomic distances around $1\mu$m where $d_{\cal B} \sim \Delta$, and finding protocols that are robust under the parallel excitation of several qubits. 
%We call this the {\em strong dipole-blockade regime} \cite{Ohmori_NP2022,Alexey_PRL2021}.

%\subsection{2-qubit structure in the strong-dipole blockade regime}

Following \cite{Sola_Nanoscale2023},
as a first approximation to obtain analytical formulas we will neglect any coupling except 
for the $|0\rangle$ and $|r\rangle$ states in each qubit. 
The complications that arise by dealing with the Stark shifts created in non-resonant two-photon transitions will be treated elsewhere.
%As a first approximation to 
We model the local effect of the field 
on each of the qubits, defining {\em geometrical factors}, $a_k$ and $b_k$, 
so the spatially and temporally dependent interaction of the laser $k$ at qubit
$\alpha$ ($\alpha = a,b$) is determined by the Rabi frequencies 
$\widetilde{\Omega}_k(\vec{r}_\alpha,t) = \alpha_k\mu_{0r} E_k(t)/\hbar
= \alpha \Omega_k(t)$.
The geometrical factors can be partially incorporated in the Franck-Condon factor 
$\mu_{0r}$ so we can assume, without loss of generality, that $a_k$ and $b_k$ 
are normalized to unity ($\sqrt{a_k^2 + b_k^2} = 1$). 
Using hybrid modes of light (structured light) 
one can control $a_k$ and $b_k$ in a wide range of values,
including negative factors.

The Hamiltonian is block-diagonal, 
${\sf H}_k^V \oplus {\sf H}_k^A \oplus {\sf H}_k^B \oplus {\sf H}^D$,
where 
${\sf H}_k^V = -\frac{1}{2} \Omega_k(t) \left(
a_k|00\rangle \langle r0| + b_k |00\rangle
\langle 0r| + \mathrm{h.c.} \right)$
is the Hamiltonian of a $3$-level system in $V$ configuration,
acting  in the subspace of $\lbrace |00\rangle, |r0\rangle, |0r\rangle \rbrace$
states, 
${\sf H}_k^A = -\frac{1}{2} a_k\Omega_k(t) \left(  
|01\rangle\langle r1| + \mathrm{h.c.} \right)$
and 
${\sf H}_k^B = -\frac{1}{2} b_k\Omega_k(t) \left(  
|10\rangle\langle 1r| + \mathrm{h.c.} \right)$ 
are two-level Hamiltonians acting 
in the subspace of $\lbrace |01\rangle, |r1\rangle \rbrace$ and
$\lbrace |10\rangle, |1r\rangle \rbrace$ respectively.
We will refer generally to any of these subsystems with the superpscript $S$
($S = V,  A, B$).
Finally, ${\sf H}^D = 0$
is a zero Hamiltonian acting on the double-excited qubit state $|11\rangle$, decoupled from any field.
%The energy diagrams of the subsystems are shown in Fig\ref{proposal}.

Using a pulse sequence of non-overlapping pulses $\widetilde{\Omega}_k(\vec{r},t)$, 
in resonance between
the $|0\rangle$ state of the qubit and the chosen Rydberg state $|r\rangle$, 
the time-evolution operator of any of
these Hamiltonians can be solved analytically through their 
time-independent dressed states, that have zero non-adiabatic couplings \cite{shore_2011,Malinovsky_2011,Sola_AAMO2018},
$${\sf U}_T^{S} = \prod_{k=0}^{N_p-1} U^{S}_{N_p-k} \ .$$
For the $V$ subsystem, 
\begin{equation}
U^{V}_{k} = 
\left( \begin{array}{ccc}
\cos \theta^V_k & i  a_{k} \sin \theta^V_k & i b_{k} \sin \theta^V_k  \\
i a_{k} \sin \theta^V_k & a_{k}^2 \cos \theta^V_k + b_k^2 & 
a_{k}b_{k} \left[ \cos \theta^V_k - 1 \right]  \\
i b_{k} \sin \theta^V_k & a_{k}b_{k} \left[ \cos \theta^V_k - 1 \right]  & 
b_{k}^2 \cos \theta^V_k + a_k^2 \end{array} \right)  \label{UV}
\end{equation}
where the mixing angle
$$\theta^V_k = \frac{1}{2} \int_{-\infty}^{\infty}\Omega_{k}(t) dt = \frac{1}{2} A_k$$
is half the pulse area.
For the two-level subsystem $A$ and $B$, we can use the same expression for
the relevant states with
$a_k = 1, b_k = 0$, for $U_k^A$, and vice versa for $U_k^B$.
However, the mixing angles depend on the local coupling:
$\theta^A_k = a_k A_k / 2$ and $\theta_k^B = b_k A_k / 2$.
We will refer to the generalized pulse areas (GPA) as $2\theta^S_k$.
% for each pulse, in this case. 
It is convenient to encapsulate all the geometrical information in the so-called
structural vectors, defining the row vector 
${\bf e}_k^\intercal \equiv \langle {\bf e}_k| = (a_{k}, b_{k})$
(${\bf e}_k \equiv |{\bf e}_k\rangle$ is the column vector in bracket notation) 
formed by all the geometrical factors for a given pulse.

\section{Classifying gate mechanisms through quantum pathways}

The overall time-evolution operator will be the time-ordered product $U_T^{S}$. The success of the implementation of the CZ  gate depends only on the first matrix element, 
$U_{T,11}^{S}$, which must be either $1$ or $-1$ depending on the subsystem considered.
For the $V$ subsystem, we can define the ``symmetrized'' states
$|\tilde{1}_k\rangle \equiv a_k |r0\rangle + b_k |0r\rangle$ and
$|d_k\rangle \equiv b_k |r0\rangle - a_k |0r\rangle$, 
where the first receives  all the coupling with the initial state $|00\rangle$
and the second is dark. For reasons that  will be clear  in the following, we
will call  $|\tilde{0}\rangle$ to  the initial state in any subsystem ($|00\rangle,
 |01\rangle,  |10\rangle$), that is, to all the computational basis. 
To use a more compact notation,
we will also use $s_k = \sin \theta^{V}_k$ and $c_k = \cos \theta^{V}_k$.
In the transformed basis, the time-evolution operator for the $V$ system has the form,
%$U^{V}_k =$
\begin{equation}
U^{V}_k = 
 c_k \left( |\tilde{0}\rangle \langle \tilde{0}| + |\tilde{1}_k\rangle \langle \tilde{1}_k| \right) 
+  i s_k \left(  |\tilde{0}\rangle \langle \tilde{1}_k| + h.c. \right) 
+ |d_k\rangle  \langle d_k |
\label{USk}
\end{equation}
For the two-level subsystems, we can identify $|\tilde{0}\rangle$ with $|01\rangle$ and
$|\tilde{1}_k\rangle$ with $|r1\rangle$ for the $A$ subsystem, and with
$|10\rangle$ and $|1r\rangle$ for the $B$ subsystem,
such that Eq.(\ref{USk}) has the same form in all subsystems, changing the mixing angles
$\theta^V_k$  for their respective values, $\theta_k^S$, and removing the
dark sector ($|d\rangle\langle  d|$) from the matrix when appropriate.
%For the coupled states, identifying $|\tilde{0}\rangle$ with $|01\rangle$ and
%$|\tilde{1}_k\rangle$ with $|r1\rangle$ for the $A$ subsystem, and with
%$|10\rangle$ and $|1r\rangle$ for the $B$ subsystem, %respectively,
%Eq.(\ref{USk}) has the same form in all subsystems,  changing the mixing angles
%$\theta^V_k$  for their respective values,  $\theta_k^S$.
We can then obtain closed expressions that are valid for all the subsystems and,
in fact, can be  asily generalized for $N$ qubit systems.

For a single-pulse ``sequence'', the only term that connects $|\tilde{0}\rangle$
at initial time with the same state at final time, is
\begin{equation}
U^{S}_{T,11}   \equiv \langle \tilde{0} | U^{S}_1 | \tilde{0} \rangle  = c_1
\end{equation}
This mechanism implies population return and requires $\theta_1^{S}$ to be odd
multiples of $\pi$, so the GPA %generalized pulse areas 
must be even, for all $S$ subsystems. 

For two-pulse sequences, 
\begin{eqnarray}
U^{S}_{T,11} = 
\langle \tilde{0} | U^{S}_2 | \tilde{0}\rangle \langle \tilde{0} | U^{S}_1 | \tilde{0}\rangle
\nonumber
\\ + \langle \tilde{0} | U^{S}_2 | \tilde{1}_2 \rangle \langle \tilde{1}_2 | \tilde{1}_1 \rangle
\langle \tilde{1}_1 | U^{S}_1 | \tilde{0}\rangle = u^{S}_0 + u^{S}_1 \ ,
\label{U2}
\end{eqnarray}
so $U^{S}_{T,11} = c_2 c_1 - \langle {\bf e}_2 | {\bf e}_1 \rangle s_2 s_1$.
where 
$\langle \tilde{1}_2 | \tilde{1}_1 \rangle = \langle {\bf e}_2 | {\bf e}_1 \rangle$ 
is the scalar product of the two structural vectors.
When %the set of coefficients for the second pulse in the sequence changes,
%the structural vector of the second pulse differs from that of the first pulse
%such that the spatial properties are changed,
the spatial properties of the second pulse differ from those of the first pulse,
$|{\bf e}_2\rangle \ne |{\bf e}_1\rangle$ and
$|\tilde{1}_1\rangle$ will overlap with $|\tilde{1}_2\rangle$ and $|d_2\rangle$.
The population can be spread over all the excited states.
%$b_{2} \ne b_{1}$, 
%$|\tilde{1}_2\rangle$ will overlap with several (or all) of the $n$ excited states in the manifold defined with respect to the new rotation along a different basis in the Hilbert  space characterized by $\widetilde{\Omega}_2(\vec{r},t)$.
%However, $\langle \tilde{0}|\tilde{1}_{k}\rangle = 0$ always.
%$\langle \tilde{0}|\tilde{1}_k\rangle = 0$ always.
In Eq.(\ref{U2}),
$u^{S}_0$ implies again the same mechanism of population transfer
where each pulse has even generalized area and induces full population return,
whereas $u_1^{S}$ provides population return to $|\tilde{0}\rangle$
after the first pulse populates $|\tilde{1}_1\rangle$ and the second drives the
population back. We call this a {\em one loop} diagram ($1$-loop), while $u_0^{S}$ is
a {\em zero-loop} diagram ($0$-loop).
In a loop, the GPA %generalized area 
of both pulses must be an odd multiple of $\pi$. 
Notice that from $|\tilde{1}_1\rangle$ one cannot further excite
the system because of the dipole blockade.

In addition to $u^{S}_0$ and $u^{S}_1$ there 
appears a novel term in three-pulse sequences, %$u^{S}_d =$
where the population remains in the Rydberg
state while the second pulse act on the subsystem, and before returning to the ground
state with $\Omega_3(t)$,
%\begin{eqnarray}
%\sum_{\displaystyle j}^n 
%\langle \tilde{0} | U^{S}_3 | \tilde{1}_3 \rangle 
%\langle \tilde{1}_3 | \tilde{1}_2^j \rangle 
%\langle \tilde{1}_2^j | U^{S}_2 | \tilde{1}_2^j \rangle
%\langle \tilde{1}_2^j | \tilde{1}_1 \rangle 
%\langle \tilde{1}_1 | U^{S}_1 | \tilde{0} \rangle
%\nonumber \\
%\begin{eqnarray}
$$u^{S}_d  = \langle \tilde{0} | U^{S}_3 | \tilde{1}_3 \rangle  \langle \tilde{1}_3 |  \tilde{1}_2 \rangle
\langle  \tilde{1}_2 | U^{S}_2 | \tilde{1}_2 \rangle \langle  \tilde{1}_2 | \tilde{1}_1 \rangle 
\langle \tilde{1}_1 | U^{S}_1 | \tilde{0} \rangle$$ %\nonumber \\
\vspace*{-1.0cm}
$$+ \langle \tilde{0} | U^{S}_3 | \tilde{1}_3 \rangle  \langle \tilde{1}_3 |  d_2 \rangle
\langle  d_2 |  U^{S}_2 | d_2 \rangle \langle d_2 | \tilde{1}_1 \rangle 
\langle \tilde{1}_1 | U^{S}_1 | \tilde{0} \rangle $$ % \nonumber \\
%\end{eqnarray}
\begin{equation}
 = - s_3 c_2 s_1 \langle {\bf e}_3 | {\bf e}_2 \rangle \langle {\bf e}_2 | {\bf e}_1 \rangle
-s_3 s_1  \left[ \langle {\bf e}_3 | {\bf e}_1 \rangle   
-\langle {\bf e}_3 | {\bf e}_2 \rangle \langle {\bf e}_2 | {\bf e}_1 \rangle  \right] %\hspace{0.5cm}
\end{equation}
% = - s_3 s_1 \langle {\bf e}_3 | {\bf e}_1 \rangle  - s_3 (c_2 -1) s_1 
%\langle {\bf e}_3 | {\bf e}_2 \rangle \langle {\bf e}_2 | {\bf e}_1 \rangle \hspace{0.5cm}
%\end{eqnarray}
that we call a {\em loop with delay} or $d$-loop.
The term in brackets  does not exist in the $A$ and $B$ subsystems.
%, where the population remains in the Rydberg state while the second pulse act on the subsystem, and before returning to the ground state with $\Omega_3(t)$.
%Under very symmetrical conditions, like in the SOP, one can make $\langle {\bf e}_2 | {\bf e}_1 \rangle = 0$, and cancel all contributions except for a single amplitude in $u^S_d$.

It is now possible to have $U^{S}_{T,11} \approx -1$ with more than one
dominating contribution, as two diagrams can be $-1$, while another one is $+1$.
However, in all the optimal protocols found, every amplitude of the pathways
was negative or, at most, slightly positive ($-1 \ge u_j^S \gtrsim 0.1$).

For four pulse sequences, one can show that
\begin{equation}
U^{S}_{T,11} = u^{S}_0 + u^{S}_1 + u^{S}_d + u^{S}_2
\end{equation}
%where 
%\vspace*{-2cm}
\begin{widetext}
\begin{equation}
\begin{array}{lllr}
    & u^{S}_0 & = c_4 c_3 c_2 c_1 & \mbox{ ... 0-loop}  \\ \\
    & u^{S}_1 & = - s_4 \langle {\bf e}_4 | {\bf e}_3 \rangle s_3 c_2 c_1 
- c_4 s_3 \langle {\bf e}_3 | {\bf e}_2 \rangle s_2 c_1 - c_4 c_3 s_2 
\langle {\bf e}_2 | {\bf e}_1 \rangle s_1
%\hspace{2.8cm} 
& \mbox{ ... 1-loops} \\ \\  % we use 1 instead of \mathbb{I}
& u^{S}_d & = - s_4 \langle {\bf e}_4 | 
\Bigl( 1 + ( c_3 - 1 ) |{\bf e}_3 \rangle \langle {\bf e}_3 | \Bigr) 
| {\bf e}_2 \rangle s_2 c_1 - c_4 s_3 \langle {\bf e}_3 | 
\Bigl( 1 + ( c_2 - 1 ) |{\bf e}_2 \rangle \langle {\bf e}_2 | \Bigr) 
| {\bf e}_1 \rangle s_1  %\hspace{0.5 cm} 
& \mbox{... 1-loops with delay} \\ \\
&  & 
- s_4 \langle {\bf e}_4| \Bigl( 1 + ( c_3 - 1 ) |{\bf e}_3 \rangle  | \langle {\bf e}_3 | \Bigr)
\Bigl( 1 + ( c_2 - 1 ) |{\bf e}_2 \rangle \langle {\bf e}_2 | \Bigr) 
| {\bf e}_1 \rangle s_1
%\hspace{1.5 cm} 
& \mbox{... 1-loop with double delay} \\ \\
& u^{S}_2 & = s_4 \langle {\bf e}_4 | {\bf e}_3 \rangle s_3 s_2 
\langle {\bf e}_2 | {\bf e}_1 \rangle s_1 
%\hspace{8.0 cm} 
& \mbox{... 2-loop} \\
\end{array}
\end{equation}
\end{widetext}
For the $A$ and $B$ subsystems, one must again remove the terms
$1  - |{\bf e}_k  \rangle \langle {\bf e}_k |$ from the  expressions, as they involve
population passage through the dark state.
%It is possible to find contributions with slightly positive values for the amplitudes, $u_j^S \lesssim 0.2$, but they do not characterize the nature of the mechanism.
% again the terms $1  - |{\bf e}_k  \rangle \langle {\bf e}_k |$  cancel out.
Analogous formulas can be derived for longer pulse sequences. %using, {\it e.g.} diagrammatic techniques. 
While the number of pathways increases exponentially with the number of pulses,
the mechanism of all protocols up to 5-pulse sequences can 
be roughly characterized using 0-loops, 1-loops, 2-loops, and d-loops.
%a rough characterization of the mechanism of all 
%protocols up to $5$-pulse sequences can be described using $0$-loops, 
%$1$-loops, $2$-loops, and $d$-loops.

Because each term is negative and their sum must be approximately $-1$,
we can define the variables
\begin{eqnarray}
x^{S} = u_0^{S} + u_1^{S} - u_d^{S} - u_2^{S} \nonumber \\
y^{S} = u_0^{S} + u_d^{S} - u_1^{S} - u_2^{S} 
\end{eqnarray}
such that 
any protocol is represented as a point within a square, 
referred to as the {\em $m$-square}.
Each apex of the square corresponds to a gate mechanism that relies on 
a single type of pathway (diagram). 
Collaborative mechanisms that involve the contribution of multiple diagrams 
are situated between these apexes, 
though the mapping is not entirely unambiguous.
%any protocol is represented by a point in a square, that we call
%the {\em $m$-square}, with 
%gate mechanisms relying on a single type of pathway (diagram) in each apex.
%Collaborative mechanisms that imply the contribution of several diagrams
%lie in between, although the mapping is not unambiguous.
Different collaborative mechanisms may share the same coordinates in the m-square,
especially around the center of the square, when both 
$x^{S}, y^{S} \approx 0$, which can be obtained
with equal contribution of $1$- and $d$-loops, $0$- and $2$-loops, or of all
diagrams at the same time. However, the advantage of using the m-square is that
it allows one to easily represent and classify %at least roughly, 
a mechanism, 
without fully listing the values of all the contributing diagrams.

Further on, to visualize the set of mechanisms used by the optimal protocols,
we partition each m-square into $9$ boxes and rank the mechanism as
a number $\omega^{S} \in [1,9]$ depending on the box where $(x^{S},y^{S})$ 
is located for subsystem $S$.
%n = int(real(div)/2.*(x2+1.)) + 1
Defining the floor integers (greatest integer smaller than the real number)
$\lfloor {x}^{S}\rfloor = \lfloor l (x^{S}+1)/2 \rfloor +1$,
$\lfloor {y}^{S} \rfloor = \lfloor l (y^{S}+1)/2]+1 \rfloor$ 
(where $l =3$ is the number of divisions of each m-square side,
$\lfloor {x}^{S}\rfloor, \lfloor {y}^{S}\rfloor \in [1,3]$), we call
$\omega^{S} = \lfloor {y}^{S}\rfloor + l( \lfloor {x}^{S} \rfloor-1 )$
the number that ranks the mechanism for each subsystem.
As explained in more detail in Sec.V, these numbers can be
represented in a cube, so-called $m$-cube, giving each mechanism three 
coordinates ordered as $\left(\omega^{A},\omega^{B},\omega^{V}\right)$,
which summarizes in a
simple visual way the mechanisms under which the gate
operates in each protocol. Obviously, the finer we divide the m-square 
into boxes
the more information we will be able to obtain. In this work
we will use a minimal division to characterize the mechanisms in the simplest
possible way.

\section{Optimal protocols}

\begin{figure}
\includegraphics[width=8.5cm]{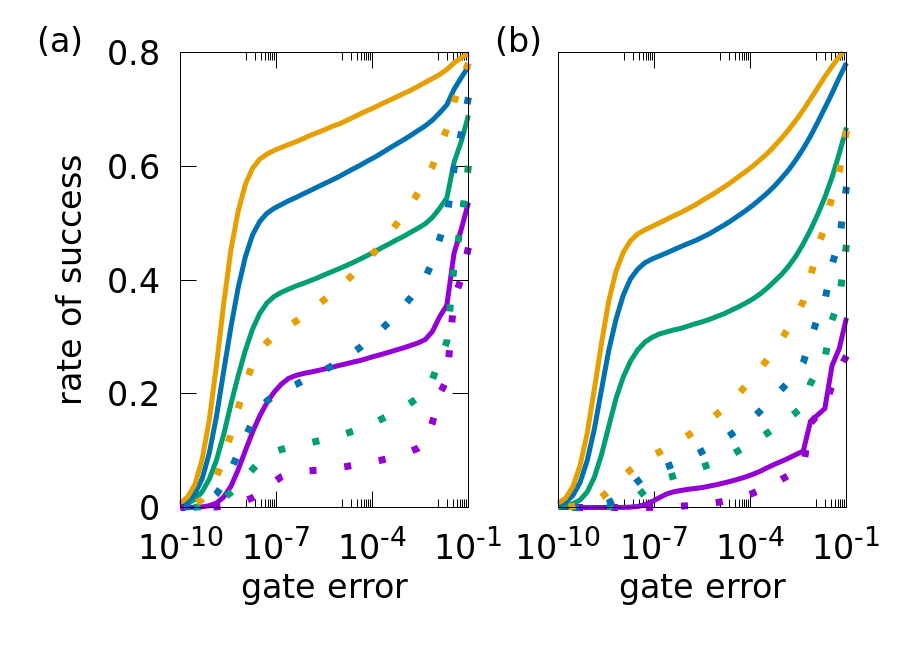}
\caption{Performance of the spatio-temporal control measured from the 
rate of success of the optimization of the gate as a function
of the error threshold, for different pulse sequences ($N_p$ from $2$ to $5$
for the lowest to highest rates)
with (a) $\sigma = 0.1$, (b) $\sigma = 0.6$. Dotted lines show the 
results obtained for p-constrained protocols.}
\label{per2q}
\end{figure}

We start by exploring the landscape of all possible optimal CZ protocols with non-overlapping
pulses in two adjacent qubits.
In this case, the optimization parameters are the effective pulse areas $A_k$
(where $k$ runs through the number $N_p$ of pulses in the sequence) and there are $2$ geometrical 
parameters per pulse. In this work $\Omega_k(t)$ are real, so the relative phase between the 
pulses is fixed as either $0$ or $\pi$.
To obtain the optimal parameters we use the Nelder and Mead simplex optimization scheme \cite{Nelder_CJ1965,Powell_AcNum1998} 
with linear constraints
starting in ${\cal N}_T = 5\cdot 10^4$ initial configurations obtained through a uniform
distribution over the parameters within some chosen range. 
The geometrical factors are constrained such that a minimum value of $|b_{k}| 
\ge \sigma$ is imposed. Protocols with smaller $\sigma$ accept
solutions where the influence of the pulse on both qubits at the same time can
be smaller, which can be related to more separated qubits.
The SOP scheme requires the orthogonality of the structural vectors, 
demanding control over the amplitude and sign of $a_k$ and $b_k$. 
In principle,  this can be  achieved using hybrid modes of light \cite{Sola_Nanoscale2023}.
But we also perform optimizations forcing the positivity of
the geometrical factors ($a_{k}, b_k \ge \sigma$) with less demanding conditions
for its experimental implementation, which we denote by $\sigma^+$ (p-restricted protocols). 

Fig.\ref{per2q} shows the {\em rate of success}, which is the 
percent of conditions ${\cal N}_\epsilon/{\cal N}_T$ 
that lead to optimal gates which perform with errors
smaller than a threshold $\epsilon$ (the fidelity being $F = 1 - \epsilon$), for sequences with 
different numbers of pulses and two values of $\sigma$: $0.1$ and $0.6$. 
The rate of success (as well as the maximum fidelity that can be achieved) 
increases with the number of pulses.
It is smaller for p-restricted protocols, particularly with $\sigma = 0.6$,
but high fidelity solutions ($\epsilon \le 10^{-7}$) can almost always be found.
%, but almost saturates already for $M = 4$. One does not need to explore larger pulse sequences.
%It is also interesting to see how the 
Optimal solutions achieve certain fidelity thresholds 
between $10^{-7}$ and $10^{-8}$ for all the different sequences, and then the probability
to find protocols with higher fidelity decays steeply.
Although the exact numbers for the rate of success may depend on the sampling of the initial parameters, the overall behavior is consistent across all sequences.
%but the overall behavior is similar. 

\begin{figure}
\includegraphics[width=7cm]{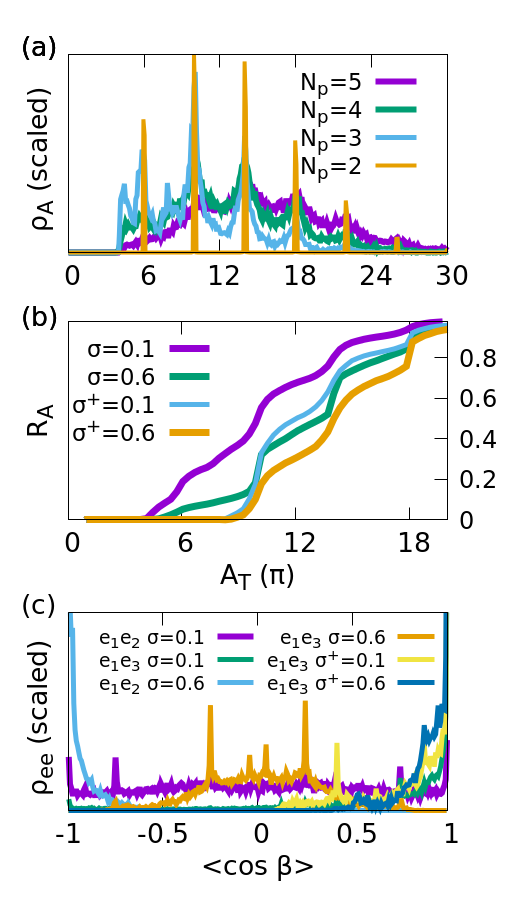}
\caption{Characterization of the parameters of the optimal protocols found by
spatio-temporal control with fidelity higher than $0.999$. 
In (a) we show the probability distribution of the protocols as a function
of the total pulse area for different pulse sequences, $\rho_A$.
In (b) we show the cumulative distribution $R_A$, as a function of the
total pulse area, for $3$-pulse sequences at different values of $\sigma$.
In (c) we show the distribution of the cosine of the angle between the structural vectors
for $3$-pulse sequences using different values of $\sigma$. 
}
\label{Atot2q}
\end{figure}

\begin{figure}
%\hspace*{-0.5cm}
\includegraphics[width=8.5cm]{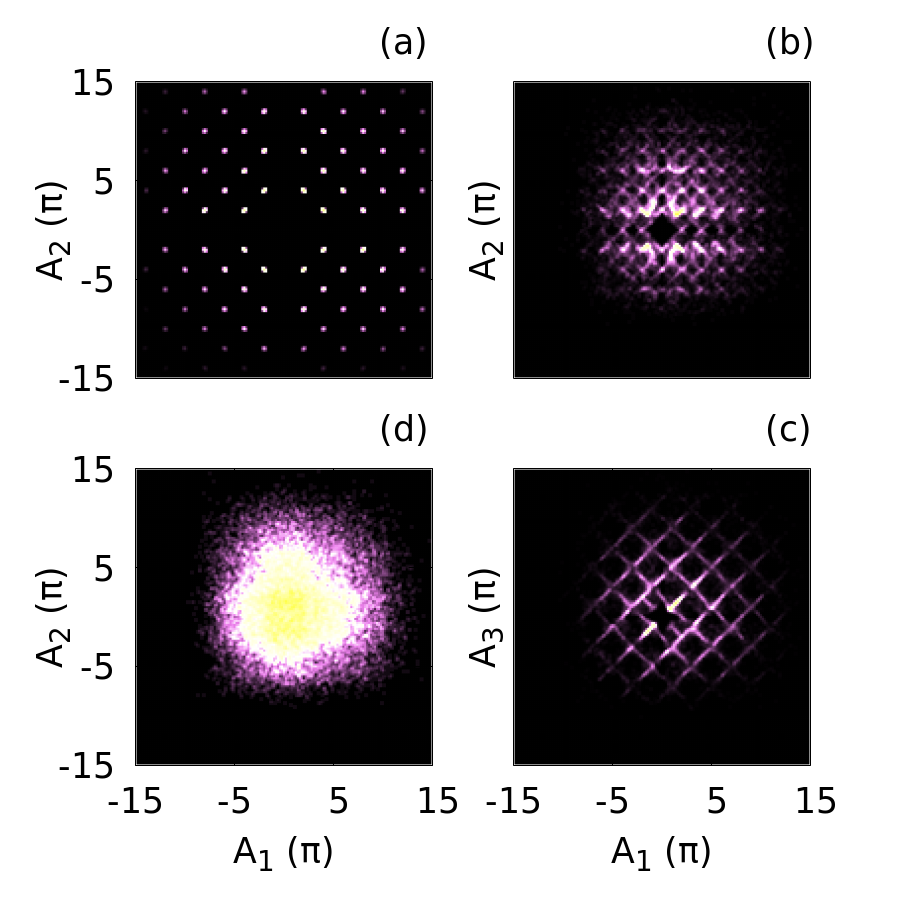}
\caption{Distribution of the pulse areas for the optimal protocols using
(a) 2 pulses, (b) and (c) 3 pulses, (d) 4 pulses. The color scale grading 
from black to red to yellow indicate the frequency of the observed values. 
In (c) we choose $A_3$ instead of $A_2$.}
\label{pdfA1A2}
\end{figure}

We can characterize the optimal solutions in parameter space or in relation to the
mechanism (dynamics) that they imply for the gate performance.
In Fig.\ref{Atot2q}(a) we represent the scaled distribution of optimal solutions
$\rho_A(A_T) = {\cal N}_A / {\cal N}_\epsilon$ as a function of the total pulse area,
where ${\cal N}_\epsilon$ is the total number of solutions 
with an error smaller than $\epsilon = 10^{-3}$, and 
${\cal N}_A$ is the subset of those solutions with a total area in the vicinity of
$A_T = \sum_k | A_k |$ (within an interval $\Delta A = 0.05\pi$). 
The results are shown for different pulse sequences with $\sigma = 0.1$.
Two-pulse sequences constrain all possible optimal solutions such that
$A_T/\pi = 6 + 4l$, $l \in \mathbb{Z}$.
The structural vectors must be completely aligned, $\bf{e}_2 = \bf{e}_1$. 
The effect of the constraints shows up in $A_T$, but also in strong correlations
in the areas of the two pulses, as shown in Fig.\ref{pdfA1A2}(a),
 where we represent
 the fraction of solutions %with error $\epsilon$ smaller than $10^{-3}$
as a function of $A_1$ and $A_2$ for the $2$-pulse sequence.
The pulse areas must alternate: one following $4l+2$, the other $4l^\prime+4$, 
($l, l^\prime \in \mathbb{Z}$).

%\begin{figure}
%\includegraphics[width=7.0cm]{minarea.png}
%\caption{Minimal total pulse area obtained for optimal protocols with a given fidelity error for pulse sequences with $N_p = 3$ (red) $4$ (green) and $5$ (violet). 
%Solid lines represent results obtained by minimizing both the fidelity error and the total pulse area, whereas dashed lines are the results obtained using the usual algorithm, restricting large area solutions.}
%\label{minarea}
%\end{figure}

Adding an extra pulse weakens the constraints, so intermediate values of $A_T$
become possible. A minimum total pulse area of $4\pi$ is necessary for high
fidelity, and particularly for $N_p=3$ one can observe maxima at $A_T = 4\pi$ and $8 \pi$.
These protocols coincide with the pulse areas in the JP \cite{Jaksch_PRL2000}
and in the SOP \cite{Sola_Nanoscale2023}.
For this set of solutions, $\langle {\bf e}_1 | {\bf e}_{2} \rangle = 
\langle {\bf e}_2 | {\bf e}_3 \rangle = 0$,  while ${\bf e}_1 = {\bf e}_3$. 
However, among the set of all possible solutions with high fidelity, 
the propensity for these values is small. 
The distribution of optimal areas changes for different values of $\sigma$ and 
in p-restricted protocols, especially in short sequences ($N_p \leq 3$). 
Fig.\ref{Atot2q}(b) shows the cumulative distribution
of protocols with a total area smaller than $A_T$,
$R_{A}(A_T) = \int_0^{A_T} \rho_A(A^\prime_T) dA^\prime_T$.
For $\sigma = 0.1$ there are protocols with minimum pulse area
$A_T \sim 4\pi$.
In contrast, p-restricted protocols need $A_T \gtrsim 9\pi$.
The step-wise behavior clearly
reveals that some values of $A_T$ are preferred, which differ 
depending on the setup.
It is possible to find protocols that use weaker fields, but at the 
expense of worsening the fidelity of the gate.
%Fig.\ref{minarea} shows the minimal total area observed for any protocol  with a given fidelity error. Only allowing errors larger than $1$\% in the fidelity one can significantly reduce the minimal pulse area beyond $4\pi$ and down to $3\pi$. The results barely depend on the number of pulses in the sequence.

In Fig.\ref{Atot2q}(c) we evaluate the correlation between
geometrical vectors for $3$-pulse sequences as a distribution of their relative orientation,
$\rho_{ee}(\cos\beta) = {\cal N}_\beta / {\cal N}_\epsilon$,
where ${\cal N}_\beta$ is the subset of solutions with a corresponding value of
$\cos\beta$ (within an interval of $0.05$) and error smaller than $10^{-3}$.
With $\sigma = 0.1$, ${\bf e}_1$ and
${\bf e}_3$ are mostly aligned, 
while ${\bf e}_2$ can take any orientation with respect to the previous vectors,
with small preferences for aligned, anti-aligned, and at angles 
$0.23\pi, 0.77\pi$,
corresponding to $\langle {\bf e}_1| {\bf e}_2\rangle = \pm 0.75$.
Interestingly, with larger $\sigma$, ${\bf e}_1$ and ${\bf e}_2$ tend to be anti-aligned,
while ${\bf e}_3$ is mostly oriented perpendicular to the other vectors,
with clear peaks in the distribution at $0.42\pi$ and $0.58\pi$ angles.
These signatures reveal different underlying mechanisms for the operation of the gate
that we believe correlate to 1-loop or 2-loop mechanisms, as we comment in Sec. IV.

Fig.\ref{pdfA1A2}(b) and (c) show that for $3$-pulse sequences one can still
find correlations among the pulse areas, especially between $A_1$ and $A_3$
(subfigure c). 
However, for pulse sequences with four or more pulses, 
almost any value of $A_T$ larger than $4\pi$ is possible,
although values of $A_T/\pi = 4l+6$ (not $4l+4$) and aligned or anti-aligned structural vectors (not orthogonal) are still
slightly preferred within the set of higher fidelity protocols.
The decay at larger values of $A_T$ observed in Fig.\ref{pdfA1A2} is artificial, 
due to the imposed range in the sampling of initial parameters.
The possible pairwise correlations between the geometrical factors or between other parameters increase 
with the square of the number of pulses, but at most, weak correlations are
observed in the set of all optimal protocols. 
To analyze the behavior of the different optimal protocols we resort now 
to the mechanism analysis introduced in Sec.III. By constraining the
protocols to obey particular mechanisms, we will show %in the following section, we will observe
that clearer correlations can be inferred between the optimal parameters.

\section{Mechanism analysis}

Optimal protocols based on $2$-pulse sequences are characterized by %imply
highly constrained values for the optimal areas and fully aligned structural vectors.
Mechanistically, when starting in the $|00\rangle$ state,
these protocols consist solely of pure $0$-loops, 
%hence 
resulting in $\left( x^{V},y^{V} \right) = (-1,-1)$.
When starting in either the $|01\rangle$ or $|10\rangle$ 
there can be %they can be 
pure $0$-loops and $1$-loops, 
as well as collaborative mechanisms that involve 
contributions of both loops. %such that 
In such cases, if one dominates when starting in $|01\rangle$, the opposite dominates when starting in $|10\rangle$.

\begin{figure}
\hspace*{-0.5cm}
\includegraphics[width=8.6cm]{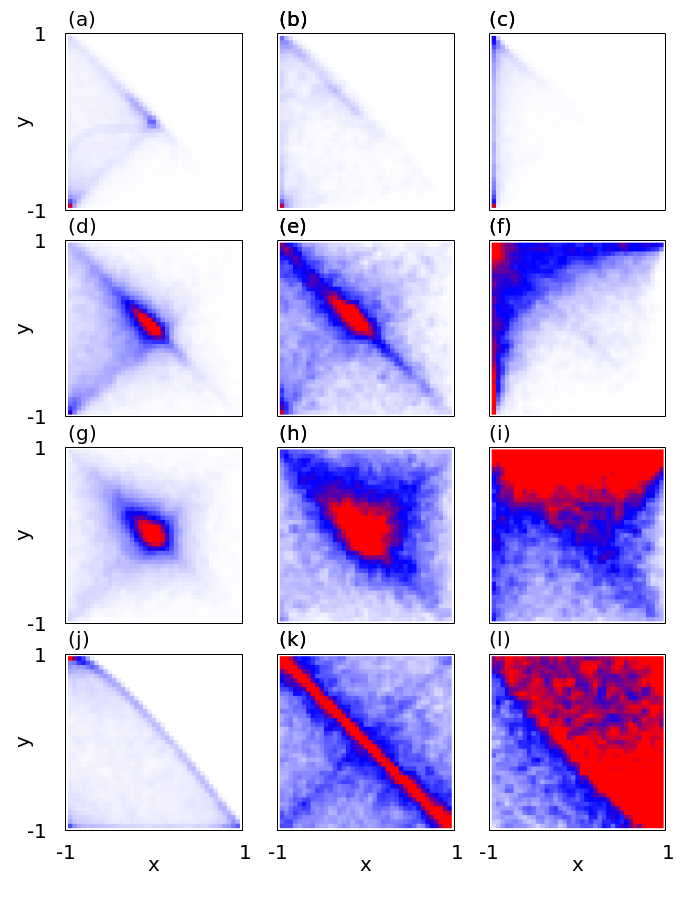}
\caption{m-square showing the dominance of different mechanisms 
in optimal protocols for the $V$ subsystem with $3$-pulse sequences (a) to (c);
$4$-pulse sequences (d) to (f); $5$-pulse sequences (g) to (i).
In the left column $\sigma = 0.1$, the center is for p-restrictive protocols
with $\sigma^+ = 0.1$ and the right column for $\sigma = 0.6$.
The last row is reserved for mechanism for the $A$ subsystem with
$N_p = 3$ (j), $4$ (k) and $5$ (l).
%For the $3$-pulse sequence the left column is reserved for 
%$\left(x^{V},y^{V}\right)$ and the right column for
%$\left(x^{A},y^{A}\right)$. Odd rows
%represent results for $\sigma = 0.1$; even rows use $\sigma = 0.6$.
%For $4$ and $5$ pulse sequences we only represent 
%$\left(x^{V},y^{V}\right)$. In this case, the right column is used
%for p-restricted mechanisms.
}
\label{square}
\end{figure}

Fig.\ref{square} gives a panorama of the mechanisms found for the $V$ subsystem. 
From top to bottom we increase the number of pulses ($N_p = 3, 4, 5$), 
and from left to right the constraints ($\sigma = 0.1$,  $\sigma^+ = 0.1$, $\sigma = 0.6$).
Except for the last row, which is dedicated to the mechanisms found in the
$A$ subsystem (which are the same as those in the $B$ subsystem) for different pulse
sequences ($N_p = 3, 4, 5$ from left to right).
The frequency of solutions is normalized to the peak value of the distribution, hence a higher
density of colors implies a broader set of mechanisms for the optimal protocols.

The most obvious conclusion is the wider choice of mechanisms (and of frequent mechanisms)
that shows up with the number of pulses or the  strength of the constraints.
Focusing on the similarities, for fixed $\sigma$ the m-squares tend to increase the
density of solutions towards the center and towards 2-loops, as the number of pulses increases.

For $3$-pulse sequences, the m-square for the $V$ subsystem with $\sigma = 0.1$
[Fig.\ref{square}(a)] shows most mechanisms lying in a triangle, involving mostly pure 0-loops 
(the dominant mechanism) and collaborative mechanisms mostly at the center of the square. 
Pure 1-loop mechanisms are very infrequent, but become more important for $\sigma^+ = 0.1$
and especially so for $\sigma = 0.6$. In the latter case, the collaborative mechanisms
involve mainly 0-loops and 1-loops, rather than d-loops.  
If we confine the mechanism analysis to the set of protocols with
smaller pulse areas ($A_T < 6\pi$ or $A_T < 10\pi$ for p-restricted protocols,
results not shown in Fig.\ref{square}) we observe the same tendency: 
a bigger contribution of d-loops in collaborative mechanisms, as pure mechanisms cease to appear.

The m-square for $4$-pulse sequences is a colorful version of the $3$-pulse case,
which brighter features towards the center.
%where collaborative mechanisms become the most prevalent. 
While 2-loops are not possible in $3$-pulse sequences, they are available in $4$ and $5$-pulse protocols
and become more important as $\sigma$ or the number of pulses increase.

\begin{figure}
%\hspace*{-0.5cm}
\includegraphics[width=8.6cm]{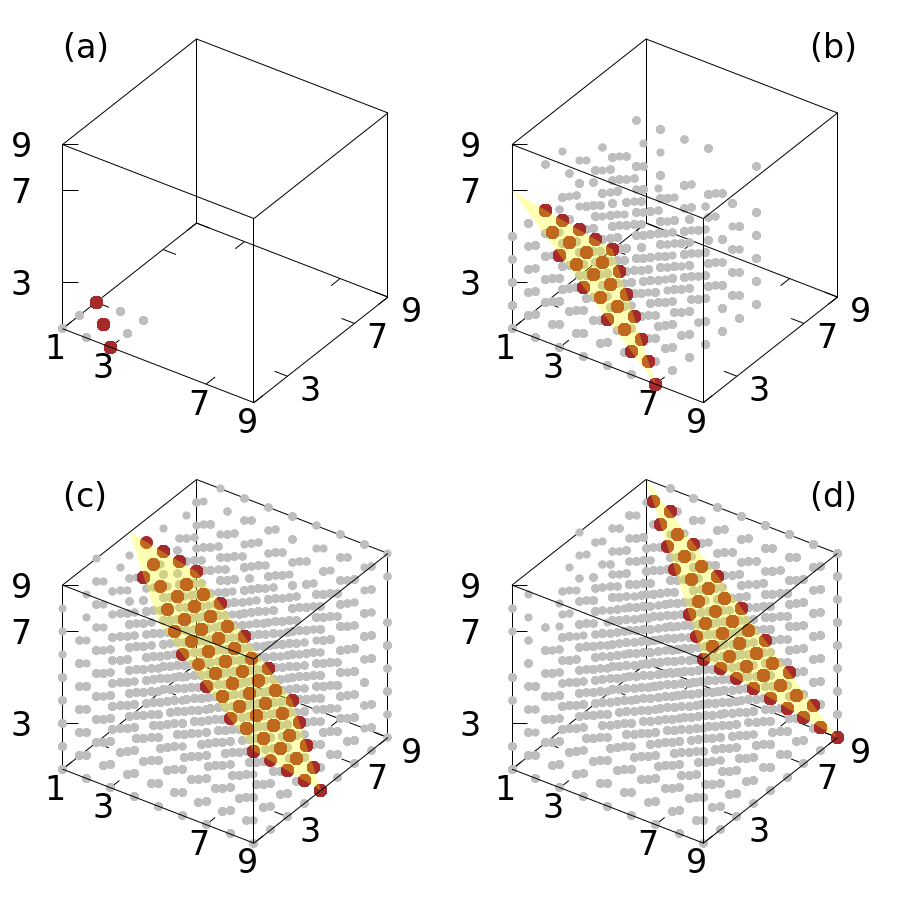} 
\caption{m-cube showing the most frequent mechanisms (brown circles) and all observed
mechanisms (gray) used by the optimal protocols with (a) 2-pulses, (b) 3-pulses,
(c) 4-pulses and (d) 5-pulse sequences. The most frequent mechanisms occupy
planes, shown in yellow.}
\label{cube}
\end{figure}

By symmetry, the m-square for the $A$ and $B$ subsystems is always the
same and typically displays a more variety of viable mechanisms than in the
$V$ subsystem, with prevalent  mechanisms along the diagonal (1-loops to
d-loops and their combinations). Pure 0-loops are still possible, but their presence is
mostly reduced to 3-pulse sequences. As $N_p$ increases, d-loops and  2-loops
become more important. For 5-pulse sequences the prevalent
mechanisms practically occupy all the upper triangle of the m-square. 
On the other hand, the diagrams are qualitatively similar regardless of $\sigma$.

But the $(x^{S}, y^{S})$ values of the different
subsystems are not independent.
To better visualize this information, we partitioned the $x$ and $y$ 
m-square for each subsystem in $9$ boxes and %we rank each mechanism by an integer number 
assigned an integer value $\omega^{S} \in [1, 9]$ 
to each mechanism based on the location of the $(x^{S}, y^{S})$ coordinates.
%depending on the location of the $(x^{S}, y^{S})$ coordinates.
%\footnote{Obviously, a finer division of the square in more cases is possible
%and gives more information, but increases the complexity of the analysis.}.
%(Obviously, a finer division of the square in more cases is possible  and gives more information, but increases the complexity of the analysis.)
Hence, for each system, pure or dominant 0-loops correspond to $\omega = 1$, 1-loops to $\omega=3$, d-loops to $\omega=7$, and 2-loops to $\omega=9$.
%In between, $\omega$ ranks collaborative mechanisms among the closest 
%pure mechanisms, or possibly fully collaborative mechanisms ($\omega = 5$). 
Collaborative mechanisms rank between the closest pure mechanisms, or possibly fully collaborative 
mechanisms ($\omega = 5$). 
We choose the $xy$ plane to represent $\omega^{V}$,  
the $yz$ plane for $\omega^{A}$ and the $xz$ plane for $\omega^{B}$.
The three values of $\omega$ characterize a point in the so-called {\em m-cube},
which is shown in Fig.\ref{cube} for the different pulse sequences with
$\sigma = 0.1$. The color indicates the probability of finding such a mechanism
among the set of optimal protocols with an error smaller than $10^{-3}$.
As a reference, Jaksch protocol is a pure d-loop for the $V$ subsystem,
a pure 0-loop for the $A$ subsystem, and a pure 1-loop for the $B$ subsystem,
occupying the $(1,3,7)$ point in the m-cube.

Fig.\ref{cube}(a) shows that all mechanisms for the $2$-pulse sequences 
lie in the $xy$ plane ($\omega^{V}=1$), and most of them are in the diagonal, 
that is, $\omega^{A}+\omega^{B} = 4$. Hence, whenever the gate performs
as a 0-loop in $A$, it works as a 1-loop in $B$ and vice versa.
The same correlation over $\omega^{A}$ and $\omega^{B}$
is observed in $3$-pulse sequences, but in a weaker form.
Now the majority of the mechanisms show up with $\omega^{S} \in [1,3]$ 
or $\omega^{S} = 7$, especially in $\omega^{V}$.
%The overall result implies a somehow %scattered distribution of dominant mechanisms in the m-cube,
%condensed distribution of dominant mechanisms 
%when $\omega^{V}+\omega^{A} + \omega^{B} \leq 9$.
The m-cube looks similar when we constrain the analysis to high-fidelity
protocols (error smaller than $10^{-7}$), so the decay in the rate of
success of the algorithms (see Fig.\ref{per2q}) has no clear implications from
the mechanistic point of view.

While the center of the m-cube is always filled with mechanisms, almost all
mechanisms are used as the number of pulses increases.
Interestingly, the preferred mechanisms lie on a single plane (shaded in yellow in Fig.\ref{cube}).
The value of $\omega_T$, which is the sum of the three 
$\omega$ values ($\omega_T = \omega^{V} + \omega^{A} + \omega^{B}$), 
is equal to $9$ for $3$-pulse sequences, $15$ for $4$-pulse sequences and $21$ 
for $5$-pulse sequences, when $\sigma = 0.1$.
This implies a surprising symmetry where the preferred optimal protocols use 
the same mechanisms regardless of the sub-system where it is applied, as the
role of $\omega^{S}$ can be interchanged between the different subsystems.
Large values of $\omega_T$ mostly correspond to favoring 2-loops over d-loops,
and d-loops over 1-loops in the colloborative mechanisms, as one moves from
$3$ to $5$-pulse sequences.
%Since $\omega_T$ is odd, 2 pure + 1 collaborative, or 3 collaborative but "even" (2, 4, 6, 8)
Similar or slightly lower values are observed for larger $\sigma$.

%For $4$ and $5$-pulse sequences all the preferred mechanisms in the m-cube lie on a single plane (shaded in yellow in Fig.\ref{cube}).The value of $\omega_T$, which is the sum of the three $\omega$ values ($\omega_T = \omega^{V} + \omega^{A} + \omega^{B}$), is equal to $17$ for $4$-pulse sequences and $15$ for $5$-pulse sequences when $\sigma = 0.1$.
%Slower or similar values are observed for larger $\sigma$. Large values of $\omega_T$ mostly correspond to favoring d-loops over 1-loops (in the m-square diagonal) in $4$-pulse sequences, and collaborative mechanisms over 1-loops in $5$-pulse sequences. Interestingly, the plane is filled for the $5$-pulse sequence, and  almost filled in the $4$-pulse sequence (where 2-loop mechanisms in $V$ are mostly absent).
%This implies a surprising symmetry where the preferred optimal protocols use the same mechanisms regardless of the sub-system  where it is applied, as the role of $\omega^{S}$ can be interchanged between the different subsystems. In fact, even in 3-pulse sequences, all dominant mechanisms can be included in just three planes with $\omega_T = 3, 5, 7$.

\begin{figure}
\includegraphics[width=8cm]{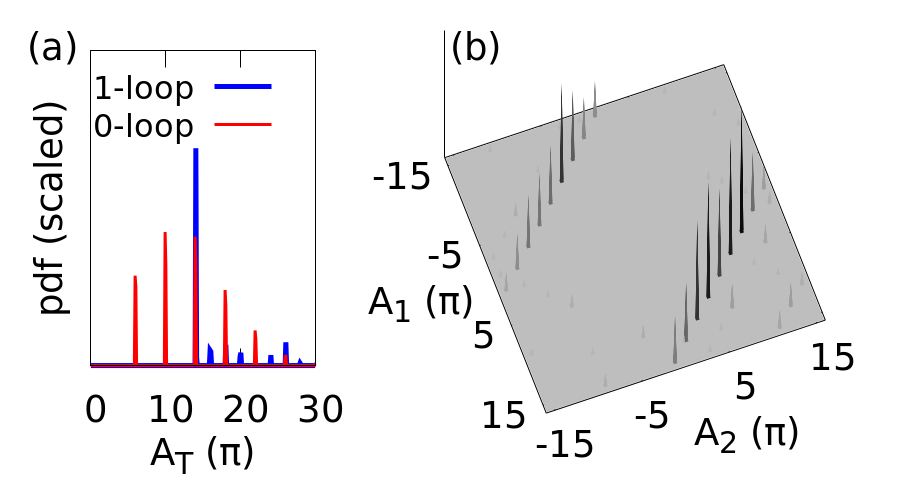}
\caption{Distribution of the pulse areas of the optimal protocols
with error smaller than $0.01$ obtained by optimizing $2$-pulse sequences
that use only pure 0-loop or 1-loop mechanisms in the $V$ subsystem. 
%starting from the $|00\rangle$ state. 
In (b) we show the correlation between the pulse areas in
1-loop mechanisms. The correlation for the 0-loop mechanism is shown in 
Fig.\ref{pdfA1A2}.}
\label{A1A2-1l}
\end{figure}

\begin{figure}
\includegraphics[width=5.5cm]{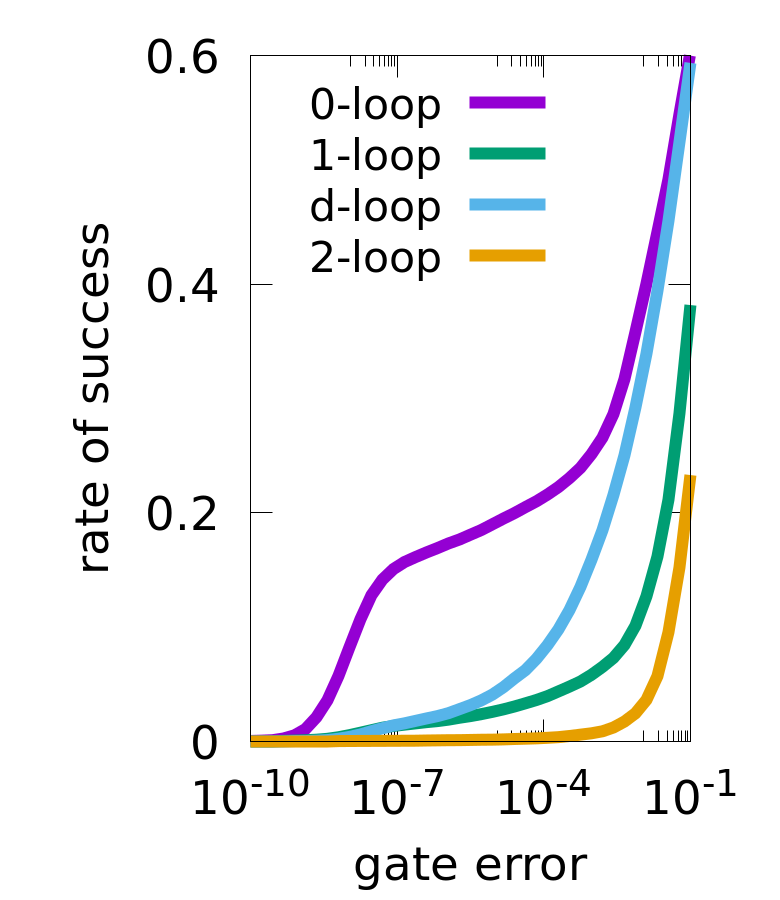}
\caption{Rate of success as a function of the threshold error,
using $4$-pulse sequences to optimize the gate following 
pre-determined mechanisms.}
\label{permech}
\end{figure}

\section{Mechanism-guided optimization}

Although the rate of success for finding optimal protocols and the
mechanisms they imply increase with the number of pulses, 
%\textcolor{red}{to thepoint of almost saturating the m-square,?????} 
it is interesting to evaluate
whether there are other optimal protocols that are not being found 
by the optimization algorithm. 
The density of solutions in parameter space suggests 
that the algorithm 
tends to find protocols that are close to the initial conditions.
%The density of solutions in parameter space seems to indicate
%that the algorithm finds protocols not too far from the starting initial
%conditions. 
%Starting 
Repeating the optimization from different sets of 
initial conditions produces similar results.
%does not change the overall results.
Very symmetric protocols occupy a negligible volume in parameter space and
typically have lower fidelities,
so one needs to impose the symmetries as restrictions in the optimization
algorithm to find them \cite{Sola_Nanoscale2023}. 

In this work, we follow a different procedure to  %and 
find optimal protocols
by maximizing the fidelity evaluated with the chosen pathways, thus
finding mechanism-driven protocols of our choice.
In the following, we constrain the optimization 
to obtain pure mechanisms in the $V$ subsystem,
while the gate may perform differently in the other subsystems, thus
selecting a specific {\em family} of mechanisms.
Using this procedure it is possible to find previously unexplored protocols
even under highly-constraining conditions.
For instance, we can find pure 1-loop protocols in 2-pulse sequences
with fidelity better than $F \ge 0.99$ (but not better than $0.999$).
Fig.\ref{A1A2-1l} shows how the pulse areas are now correlated,
following a very different pattern than in 0-loop protocols. 
The correlation between $A_1$ and $A_2$ differs and $A_T$ has different values
than before and is typically larger.
Although most protocols imply aligned structural vectors, anti-aligned
vectors are also possible.

\begin{figure}
\includegraphics[width=8.5cm]{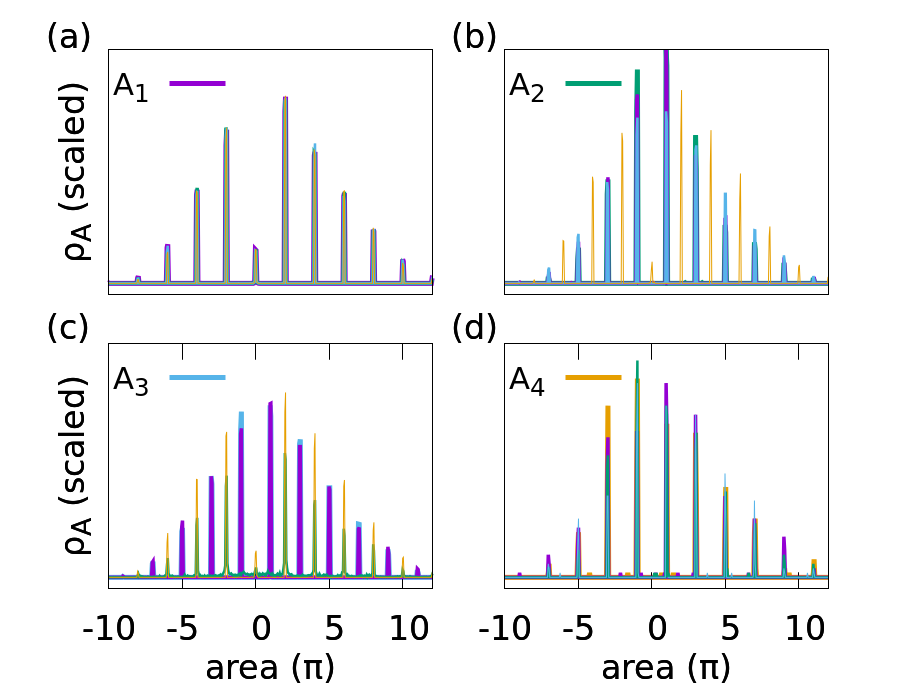}
\caption{Time-domain features of the optimization. 
Distribution of the pulse areas $A_1$ (violet), $A_2$ (green),
$A_3$ (blue), $A_4$ (orange) for the optimal protocols that 
use only (a) 0-loops, (b) 1-loops, (c) d-loops, and (d) 2-loops in 
the $V$ subsystem.}%$V^{(2,1)}$.}
%starting in the $|00\rangle$ state.}
\label{parsel1}
\end{figure}

\begin{figure}
\includegraphics[width=8.5cm]{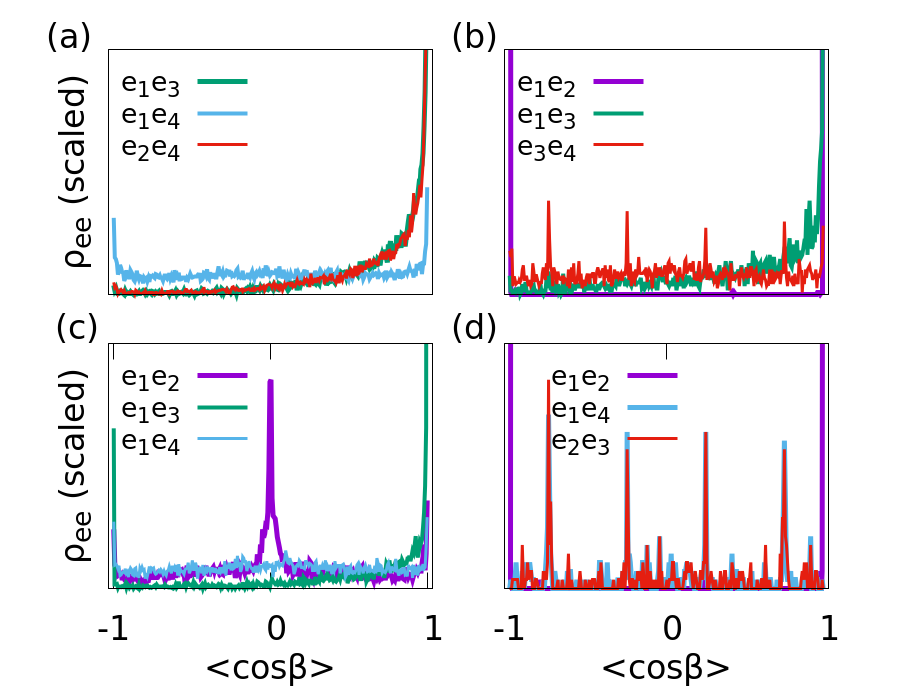}
\caption{Spatial-domain features of the optimization. 
Distribution of the cosine of the angles between structural vectors
for the optimal protocols that only use
(a) 0-loops, (b) 1-loops, (c) d-loops, and (d) 2-loops in the
$V$ subsystem.} %$V^{(2,1)}$.}
%starting in the $|00\rangle$ state.}
\label{parsel2}
\end{figure}

In Fig.\ref{permech} we show the rate of success of the
optimization at selecting protocols %with pure mechanisms in $V^{(2,1)}$ %pure-mechanism protocols 
%for processes starting in the $|00\rangle$ state 
in $4$-pulse sequences.
As expected from our previous analysis without mechanism selection, 
pure 0-loops are easy to find even in gates performing at high-fidelity.
Lowest fidelity protocols are achieved by pure 2-loop protocols, hence their
absence in the unconstrained optimizations.

Fig.\ref{parsel1} shows the distribution of pulse areas for the different
mechanisms, and Fig.\ref{parsel2} some representative correlations between structural vectors.
Protocols with pure-mechanisms constrain the pulse areas such that %to particular values:
in 0-loops and 2-loops all pulses have the same areas; in 1-loops, the areas
of the fourth pulses $A_4$ differ, and in d-loops odd pulses and even pulses have different 
areas. %pulses in the sequence take different values.
The total pulse areas in 0-loop protocols follow the same rules as in the
2-pulse sequences: $A_T/\pi = 6 + 4l$. In all other cases, the
rule is $A_T/\pi = 6 + 2l$ except in d-loops, which also allow total pulse areas
smaller than $6\pi$.
In average, pure 0-loop and 2-loop protocols typically
require larger $A_T$ than 1-loop and d-loop protocols.
Collaborative mechanisms can further reduce the pulse areas.

The analysis of the vector correlations shows the following: 
In 0-loop protocols,
{\em odd} vectors (${\bf e}_1, {\bf e}_3$) as well as {\em even} vectors
(${\bf e}_2, {\bf e}_4$) are mostly aligned with each other, while odd to even
vectors show up at any possible orientation, with some preference for
aligned or anti-aligned configurations.
%although it is slightly more common to find them aligned or anti-aligned.
In 1-loop protocols, ${\bf e}_1$ and ${\bf e}_3$ are strictly aligned or anti-aligned 
to ${\bf e}_2$, while they are mostly aligned to each other;
${\bf e}_4$ appears at all possible orientations but the distribution shows peaks at
$0.23\pi, 0.42\pi, 0.58\pi$ and $0.77\pi$, corresponding to 
$\langle {\bf e}_4| {\bf e}_j\rangle = \pm 0.25, \pm 0.75$.
In d-loop protocols, ${\bf e}_1$ and ${\bf e}_3$ are both strictly aligned or
anti-aligned to each other, and perpendicular to ${\bf e}_2$, while ${\bf e}_4$
takes all possible orientations.
Finally, in 2-loops ${\bf e}_1$ is aligned or anti-aligned to ${\bf e}_2$, and
${\bf e}_3$ to ${\bf e}_4$, while ${\bf e}_3$ and ${\bf e}_4$ show at
$0.23\pi, 0.42\pi, 0.58\pi$ and $0.77\pi$ angles with respect to both ${\bf e}_1$
and ${\bf e}_2$.

\section{Summary and Conclusions}

%We have analyzed optimal sequences of non-overlapping pulses %
%that allow for %with 
%control over the spatial degrees of freedom in trapped neutral atoms.
%. %in ideal conditions.
%We have explored 
%exploring the space of optimal protocols 
%These models explore the space of optimal protocols for 
%implementing % that implement 
%CZ entangling gates in systems of two  
%non-independent qubits, with high fidelity.
Using quantum control tools, we have explored the space of optimal protocols for 
implementing CZ entangling gates in systems of two non-independent qubits, with high fidelity.
%These qubits can form a denser quantum media allowing to boost the dipole blockade, so that the gates could in principle operate in the nanosecond regime.
Studying the rate of success of the optimal control algorithm as a function of the gate error 
for different pulse sequences under different constraints, 
we have evaluated the impact of the proximity of the atoms. 
High fidelity protocols can be found already with $2$-pulse sequences
in highly interdependent qubits, where each field acts
strongly on both qubits at the same time. % ($\sigma = 0.6$).
%, even imposing positive fields everywhere (p-constrained protocols).
However, the density of solutions decreases as the qubits approach 
each other, and in
a more pronounced way if the fields are forced to be positive everywhere. % (p-constrained protocols).
%for p-constrained protocols.
%It is even possible to find single-pulse protocols, albeit only for low-fidelity gates.
The minimal pulse areas necessary to implement the protocols 
also increase with the constraints.
%, starting around $A_T = \sum_k A_k \sim 3.9\pi$, %also 
%increase with $\sigma$ and particularly so in p-constrained protocols.
%The parameters of the pulses are strongly correlated.

To characterize the protocols up to $5$-pulse sequences, 
we have proposed a mechanism analysis based on pathways that connect the initial 
computational state of the qubit with the final state, in terms of 0-, 1-, d-, and 2-loops, 
represented on a square.
We have approximately ranked the solutions 
in terms of pure mechanisms or their combinations,
characterizing each protocol by a point in a cube.
Finally, we have developed optimization algorithms that select protocols that operate
under chosen mechanisms.
%
%The numerical results showed that a minimum of $3.9\pi$ of total pulse area is needed in the pulse energies for the CZ gate to perform at high fidelity, although the areas could drop to $3\pi$ allowing a $10$\% error.
% symmetric protocols are not found except imposing restrictions
%
%We have found strong correlations in the parameters of all 2-pulse optimal protocols, such that only very precise values of the pulse areas are possible, starting with $A_T = 6\pi$. Both pulses must have the same spatial features.

All protocols in 2-pulse sequences
%All these protocols  imply 
require a 0-loop mechanism for the dynamics starting in $|00\rangle$
(for the dynamics starting  in $|01\rangle$ or $|10\rangle$ the mechanism can be
a 0 or 1-loop or its superposition).
But lower-fidelity protocols can be found forcing a 1-loop  mechanism in $|00\rangle$,
at the expense of needing larger  pulse areas.
The correlations in the parameters are %mostly hidden 
not obvious for longer pulse sequences but can be found by
imposing mechanism constraints. 
For instance, $4$-pulse sequences that implement pure mechanisms inherit much 
of the structure of $2$-pulse sequences.
Some mechanisms involve preferred orientations in the structural vectors
and probably reveal interesting Hamiltonian structures that are
exploited in the gate dynamics, in the same  way that the SOP used a dark state \cite{Sola_Nanoscale2023}.
%Typically, protocols that work under pure $0$-loop and $2$-loop mechanisms require stronger pulses than those that work using $1$-loop, $d$-loop  and collaborative mechanisms.

While for large pulse sequences almost any possible mechanisms is used by different optimal
protocols,
the set of preferred mechanisms lie on a single plane, revealing that strong
correlations also characterize the space of mechanisms.
These correlations are such that  for any dominant mechanism,
by interchanging the type of controlled dynamics starting in any
computational basis of the qubit (except the uncoupled $|11\rangle$ state),
one can find an alternative dominant optimal protocol.
As the number of pulses increases, or the constraints become stronger, 
collaborative mechanisms are favored where the largest contributions move
from 0-loops to 1-loops, from 1-loops to d-loops, and from d-loops to 2-loops. 
The mechanisms for the dynamics starting from the $|01\rangle$ or the $|10\rangle$
states are typically more varied than those starting from $|00\rangle$, but
less dependent on the constraints.

%The preferred mechanisms also depend on the allowed proximity of the qubits and the existence of additional constraints in the field structure, such as positive geometrical factors.

%Fast gates, operating in the nanosecond, are inherently more robust to decoherence effects. Preliminary studies show that the gates are also relatively robust to the thermal  motion of the atoms, as well as fluctuations in the pulse intensities. Further studies are needed  to assess the effect of nonlinear effects in the Hamiltonian, not included in our models, as well as the practical limitations in using structural light.

From the theoretical point of view, our study offers a novel methodology to map
and characterize a dense space of optimal protocols of general validity
for quantum computing, regardless of the gate or specific platform.
To date, most proposed quantum protocols were based on human ingenuity,
forcing very restricting sets of parameters.
These highly symmetrical protocols %with  strong restrictions in the parameters,
typically implied gates that operated with pure mechanisms.
However, in the full space of mechanisms, such protocols can only be found
by guiding the search, biasing the optimization algorithm.
At  the expense of increasing the complexity of the system, controlling the spatial
properties of the laser  beams working  with structured light,
%the tools  that  we used in this  work have shown 
we have shown in this work that the landscape of protocols is much richer
than expected, and exploring this landscape offers %offering 
great flexibility for the experimental implementation.
% JUST THE SURFACE OF IT: MACHINE LEARNING  what parameters  are important?  *** ETC
%even if the spatio-temporal control with hybrid modes of light suggested in Nanoscale for a very specific protocol, or the new implementations offered in this manuscript for any implementation, were found to be too challenging or inconvenient to implement experimentally, the theoretical analysis provided in the first two points would still be valid.
%In the future, we expect that machine learning tools will reveal in a more systematic  way the

\section*{Acknowledgements}
%\begin{Acknowledgements}
This research was supported by the Quantum Computing Technology Development Program (NRF-2020M3E4A1079793). IRS thanks the BK21 program (Global Visiting Fellow) for the stay during which this project started and the support from MINECO PID2021-122796NB-I00. SS acknowledges support from the Center  for Electron Transfer funded by the Korean government(MSIT)(NRF-2021R1A5A1030054)
%\end{Acknowledgements}

\bibliography{SolaQST23.bib}
%\end{thebibliography}
%\bibliography{maintext_PRXQ.bib} %You need to replace "rsc" on this line with the name of your .bib file
%\bibliographystyle{rsc}

\end{document}